\newif\ifmulticol	\multicoltrue
\newif\ifshowgit	\showgittrue		% switches footer on/off
\newif\ifgitlocal	\gitlocaltrue		% use local file gitHeadLocal.gin
\newif\ifbiblatex	\biblatextrue		% defaults to bibtex if false
\newif\ifbibnum		\bibnumtrue 		% num => superscripts, otherwise auth date
\newif\ifbibsort	\bibsortfalse		% biblatex num sort in order of occurrence
\newif\iflineno		\linenofalse
\newif\iftoc		\tocfalse

\newif\iflucida		\lucidafalse
\newif\ifcm			\cmfalse
\newif\iflibertine	\libertinefalse		% use for arXiv when bold math needed
\newif\ifcharter	\chartertrue

\newcommand*{\secnumdepth}{0}			% sets secnumdepth in \mymaketitle
\newcommand*{\tocdepth}{2}				% sets tocdepth in \mymaketitle
\newcommand*{\reftitle}{References}		% title for references section
\newcommand*{\mytitleskip}{-0.2in}		% change skip after title

\newcommand*{\mydocfontsize}{\ifcharter11pt\else\iflibertine11pt\else10pt\fi\fi}
\newcommand*{\setcol}{\ifmulticol twocolumn\else onecolumn\fi}

\documentclass[\mydocfontsize,\setcol]{article}

\newcommand*{\pdfauthor}{Steven A.~Frank}
\newcommand*{\pdftitle}{Robustness and complexity}

\input robust.sty
\usepackage{bm}

\DeclarePairedDelimiter\abs{\lvert}{\rvert}
\DeclarePairedDelimiter\norm{\lVert}{\rVert}
\DeclarePairedDelimiter\angb{\langle}{\rangle}
\DeclarePairedDelimiter\lrb{\lbrack}{\rbrack}
\DeclarePairedDelimiter\lr{\lparen}{\rparen}
\DeclarePairedDelimiter\lrbr{\lbrace}{\rbrace}

\makeatletter
\let\oldabs\abs \def\abs{\@ifstar{\oldabs}{\oldabs*}}
\let\oldnorm\norm \def\norm{\@ifstar{\oldnorm}{\oldnorm*}}
\let\oldangb\angb \def\angb{\@ifstar{\oldangb}{\oldangb*}}
\let\oldlrb\lrb \def\lrb{\@ifstar{\oldlrb}{\oldlrb*}}
\let\oldlr\lr \def\lr{\@ifstar{\oldlr}{\oldlr*}}
\let\oldlrbr\lrbr \def\lrbr{\@ifstar{\oldlrbr}{\oldlrbr*}}
\makeatother

\newcommand*{\boldrule}{\hrule height 1.2pt}
\newcommand*{\noterule}{\medskip\boldrule\medskip}	% for notes

\hypersetup{draft}

\newcommand*{\mytitle}{\pdftitle}

\newcommand*{\myauthor}{Steven A.\ Frank}
\newcommand*{\myaddress}{Department of Ecology and Evolutionary Biology, University of California, Irvine, CA 92697--2525, USA}
\newcommand*{\mycontact}{\href{https://stevefrank.org}{https://stevefrank.org}}

\newcommand*{\mykeywords}{Evolution, system design, paradox of robustness, constructive neutral evolution, genomic complexity, deep learning}
\setabstract{-0.07}{\iftoc-2.0\else1.22\fi}{%
When a biological system robustly corrects component-level errors, the direct pressure on component performance declines. Components may become less reliable, maintain more genetic variability, or drift neutrally in design, creating the basis for new forms of organismal complexity. This article links the protection-decay dynamic to other aspects of robust and complex systems. Examples include the hourglass pattern of biological development and Doyle's hourglass architecture for robustly complex systems in engineering. The deeply and densely connected wiring architecture in biology's cellular controls and in machine learning's computational neural networks provide another link. By unifying these seemingly different aspects into a unified framework, we gain a new perspective on robust and complex systems.
}

\begin{document}

\mymaketitle

%% 1st parm is skip on left column at start of TOC, 2nd param is skip after TOC
\iftoc\mytoc{-24pt}{\newpage}\fi

\begin{quote}
The ultimate result of shielding men from the effects of folly, is to fill the world with fools. --- Herbert Spencer \autocite{spencer07essays:}
% quote on page 354
\end{quote}

\section{Introduction}

The more strongly a robust system protects itself from the failure of its components, the more the system's components will tend to decay in performance. Suppose, for example, that our bodies added another protection against cancer. Then, a breakdown in an existing protection would have less consequence because the extra protection provides an additional check against disease \autocite{frank04geneticb}.

Reduced consequence means that the direct pressure of natural selection on existing components has weakened. Less selective pressure leads to evolutionary decay. The ultimate result of shielding a system from the failure of its components is to fill the system with weakened components. I call that the paradox of robustness \autocite{frank07maladaptation}.

The logic is so simple and compelling that it must be true. But is it important? How much of evolutionary pattern and biological design arise from the paradox of robustness? 

The answers remain unclear. Part of the difficulty is that the paradox of robustness focuses too narrowly. Instead, we must think more broadly about how robustness influences the architecture of organismal design.

I build toward that broader perspective through a series of steps. The first section develops the paradox of robustness by expanding the cancer example and adding an engineering example from the history of computer hard drives and data storage. Those examples clarify how system robustness leads to component decay and to greater complexity of design.

The second section links various ideas to the paradox of robustness, particularly the theory of constructive neutral evolution \autocite{stoltzfus99on-the-possibility,stoltzfus12constructive}. The similarities and differences between these theories help to build a broader framework.

The third section reviews observed patterns of robust and complex systems. The hourglass pattern of development \autocite{shook08morphogenic,irie14the-developmental} and the hourglass pattern for the architecture of robust systems \autocite{doyle11architecture} provide interesting examples, suggesting an expanded conceptual foundation for robustness and complexity.

The final section illustrates the new theory's perspective. In machine learning, deeply and densely connected computational neural networks revolutionized artificial intelligence \autocite{goodfellow16deep}. Similarly, deeply and densely wired regulatory control architectures of cells, which may have arisen as a consequence of the paradox of robustness, could have accelerated evolutionary adaptation in the history of life \autocite{frank17puzzles5}.

\section{The paradox of robustness}

I illustrate the theory with two examples, cancer and computer hard drives.

To protect against cancer, our bodies have multiple protections. Tumors progress as those protections break down \autocite{weinberg07the-biology}. For example, several checkpoints act as brakes on the cell cycle. Knockouts of those brakes allow continuous cell division, favoring tumor growth. At the cellular level, damage often induces cell suicide, culling aberrant and potentially precancerous cells. Knockout of the normal apoptotic suicide program promotes cancer.

Different tissues in our bodies seem to have different numbers of protections against cancer \autocite{frank07dynamics}. The same tissue in different organisms seems to have different numbers of protections \autocite{nunney20determining}. In other words, the amount of protection seems to be evolutionarily labile.

That evolutionary lability leads to a thought experiment \autocite{frank04geneticb}. What happens when an extra protection gets added? Initially, the system more robustly protects against perturbations that cause disease because there is one more factor that limits the spread of a tumor. That enhanced system robustness also changes the pressure of natural selection acting directly on the protective components that were already present. For example, losing a brake on the cell cycle is less important if there is now another apoptotic mechanism that can detect such damage and kill the cell.

Weakened selective pressure enhances the spread of mutations \autocite{visser03perspective:} and the heritability of disease \autocite{frank23robustness}. The reduced benefit provided by a particular component may also cause that component to decay evolutionarily to a less costly, lower performing, and sloppier state \autocite{frank07maladaptation}. 

As the components decay, the newly added protection becomes evolutionarily irreversible  \autocite{frank07maladaptation}. Removing that protection now exposes the lower performing components without the additional protection. The system would perform poorly. Thus, additional robust protection and the subsequent evolutionary relaxation of the prior components lead to an irreversible increase in complexity. 

This relation between enhanced system robustness and component decay follows simple logic. As the system becomes better at protecting against failure, fluctuations and sloppiness in component performance matter less. Enhanced system robustness associates with decaying component performance. That logic applies broadly, to any system evolving with respect to performance.

The logic is so simple and general that it would seem to be a fundamental principle of evolutionary design. However, it is challenging to find compelling examples in biology \autocite{lynch12evolutionary}. One difficulty is that we cannot easily see the steps by which this evolutionary process occurs. We would need evidence for the origin of a new mechanism that enhances robustness at the system level. We would then need to trace the history over which various components of the system decay in performance.

In searching for examples that illustrate the steps of increasing system robustness and decaying component performance, the best case that I could find comes from the engineering history of computer hard drives and data storage.

Many years ago, a small hard drive was expensive. Part of the expense arose from the need to make the drives reliable, with low failure rates and low error rates. Drive failures cause catastrophic loss of data. Data errors cause loss of confidence, eventually rendering the data useless.

The primary approach to data storage changed over time. Instead of focusing on reliable and expensive individual drives, storage design emphasized Redundant Arrays of Inexpensive Disks, or RAID arrays \autocite{patterson88a-case}. Redundancy enhances reliability by making copies of the data. For example, copies of data may be stored on two disks. If one drive fails, the other has a full copy of the data and nothing is lost. However, making two fully redundant copies slows performance and doubles the number of drives required, increasing the cost.

To gain the benefit of redundancy and mitigate the costs, RAID arrays often use special RAID controllers that are small computers sitting above the data storage array. When the data come in to be stored, the RAID controller breaks up the data into small chunks and spreads those data chunks across the array in a partially redundant manner. More data copies enhance protection against the failure of individual drives but also increase costs and reduce performance. One can tune the redundancy to achieve particular goals of reliability, cost, and performance. Most modern mid-level and high-level computing systems use some variant of RAID data storage.

If a drive in the RAID array fails, one can pull out that drive and put in a new one without turning off the system. The RAID controller uses the redundant data on the other drives to fill the new drive with the same data held by the failed drive. The system fully recovers while continuing to run.

Here is the key point with regard to the paradox of robustness. Because a failed drive causes relatively little disruption, it is no longer so important that individual drives be engineered to high reliability at large expense. Instead, system designers choose relatively inexpensive disks that have relatively high failure rates \autocite{backblaze14}.

The robustness gained by designing reliability at the higher system level causes a shift in the marginal costs and benefits of component disk performance. The best design typically allows a decay in component disk performance, leading to a reliable system that has cheaper, lower-performing, and sloppier components.

In engineering, if we wish to redesign the system, we can throw out the current design and start over. In biology, the greater robustness achieved by adding a high-level manager above the component parts will often be irreversible because the lower-level components will evolve to depend on the higher level of control and protection.

\section{Related theories}

The cancer and RAID examples introduced the paradox of robustness. I now describe some related theories to give a sense of similar topics and to broaden the conceptual framing of the subject. I start with brief summaries of three ideas to provide some historical perspective. I then develop the theory of constructive neutral evolution in detail.

%Constructive neutral evolution is similar to the paradox of robustness but emphasizes different applications of the theory. By linking constructive neutral evolution and the paradox of robustness, we will gain a broader perspective from which we can see more clearly how the concepts apply to problems of biological design. In the following section, I use this broader conceptual perspective to discuss interesting puzzles in the evolution of development and the architecture of organismal design.

The first idea comes from Susan Lindquist. She studied cell biology systems in which one protein buffers the effects of variability in other proteins \autocite{rutherford98hsp90}. In the absence of the buffer, amino acid substitutions typically reduce protein performance. Some of those deleterious amino acid substitutions become functionally neutral in the presence of the buffer. I describe a specific example below.

The point here is that buffering causes a kind of robustness in which changes that were previously deleterious become neutral. In the presence of the buffer, those neutral variants will accumulate in the population, increasing genetic diversity. In essence, the buffered proteins decay in their performance when isolated from their system's robust protection, a kind of evolutionary relaxation in response to system robustness.

The second idea comes from the theory of neutral networks \autocite{schuster94from,wagner08neutralism}. We start with a network of interacting components within cells. Given our focus on robustness, we imagine some higher-level process that renders alternative network interactions nearly equivalent in function. The neutrality of alternative network interactions leads to evolutionary drift in those interactions.

Eventually, the interacting components may arrive at a state from which they can achieve a significantly altered way of functioning or a significantly better way to adapt to a novel environmental challenge. Put another way, the neutrality imposed by robustness leads to wide neutral exploration and subsequent novelty in design and function.

The third idea concerns the theory of fitness landscapes \autocite{fragata19evolution,malan21a-survey}. Recently, that theory has been developed most extensively in the study of viruses because one can measure genotype, phenotype, and fitness more easily than for most other organisms \autocite{delgado23viral,schuster23virus}. Fitness landscape theory has not been linked to the paradox of robustness but could be an important future development.

The paradox of robustness essentially describes how changes in system robustness tend to flatten the fitness landscape that shapes the evolution of the system's components. The flatter landscape leads to less intense selection, more variability, and altered marginal costs and benefits. Explicit analyses of those changes may provide further insight.

I now turn to the most important related theory, constructive neutral evolution. This theory originated in the 1990s in Dalhousie, Canada, predating my own work on the paradox of robustness by about ten years. The work was developed by Michael Gray, Arlin Stoltzfus, Ford Doolittle, and many others \autocite{covello93on-the-evolution,stoltzfus99on-the-possibility,gray10irremediable,stoltzfus12constructive,brunet18the-generality,munoz-gomez21constructive}. I only learned about constructive neutral evolution recently. One goal for this article is to bring the complementary insights of constructive neutral evolution and the paradox of robustness together to advance our understanding over a broader set of biological problems.

RNA editing provided the first example of this theory \autocite{covello93on-the-evolution}. Typically, DNA makes RNA makes protein. In some organisms, DNA makes RNA, the RNA sequence is altered by an editing process, and the edited RNA sequence makes protein. For example, C nucleotides in the RNA may be converted to U nucleotides. U nucleotides in RNA are analogous to T nucleotides in DNA. The C $\rightarrow$ U change means that, in the RNA, a U remains U, and a C becomes U.

In the absence of RNA editing, a DNA nucleotide G codes for an RNA C, and a DNA nucleotide A codes for an RNA U. In the presence of RNA editing in which C $\rightarrow$ U, the DNA nucleotides G and A both code for RNA U. Thus, RNA editing causes neutrality at the DNA level between G and A nucleotides, leading to drift in the frequency of those nucleotides at particular sites in the DNA sequence.

If the RNA editing process were removed, some of the G and A DNA nucleotide variants would associate with different amino acids in the protein. The majority of amino acid changes would likely be deleterious. Thus, once RNA editing is in place and the associated DNA nucleotides drift in frequency, it will often be difficult evolutionarily to remove the RNA editing process.

An evolutionary ratchet occurs. RNA editing causes neutrality at the DNA level. Drift occurs. Removal of RNA editing leads to new deleterious DNA variants, disfavoring loss of the RNA editing process. 

The general scenario leads to a ubiquitous force of genomic complexification. First, a new mechanism of genomic processing arises. That mechanism buffers variability in another process, rendering some variants neutral. Drift follows. The new buffering mechanism cannot be removed without deleterious consequences. The genomic processing system has become irreversibly more complex. Constructive neutral evolution has occurred.

Eukaryotic genomes often seem irrationally complex. Constructive neutral evolution shows how such complexity may arise nonadaptively, as a consequence of buffering or robustness mechanisms \autocite{gray10irremediable}.

A second example of constructive neutral evolution comes from Susan Lindquist's work on cellular buffering, briefly mentioned at the start of this section \autocite{rutherford98hsp90}. Lindquist worked on the heat shock protein Hsp90. This protein helps other proteins to fold correctly into functional three-dimensional structures. In the absence of Hsp90, primary amino acid sequence variants may misfold. In the presence of the Hsp90 folding chaperone, some of those sequence variants fold into approximately equivalent functional shapes.

Lindquist realized that the Hsp90 chaperone adds robustness to the system, effectively buffering amino acid variation and causing different genetic variants to be selectively neutral. Lindquist emphasized that such robustness and associated increase in genetic variation may enhance future adaptation. In a subsequently changed environment,  some of those currently neutral variants might become advantageous, allowing rapid evolutionary response to the changed environment.

Lindquist developed her ideas in the 1990s, around the same time as the theory of constructive neutral evolution first arose. The ideas are similar. However, Lindquist focused on genetic variation and future evolutionary response. By contrast, constructive neutral evolution emphasizes the complexification of cellular process. Once protein folding chaperones are in place and the buffered neutral variation follows in the primary amino acid sequences, removing the chaperone may be significantly deleterious. An essentially irreversible complexification of cellular process occurs.

Finally, in my own work on the paradox of robustness, I have emphasized that genomes are overwired \autocite{frank17puzzles5}. By that, I mean that the regulatory network of key processes seems to contain a very large number of inputs into particular functions. An engineer designing such a control system would not create such a complexly wired network that is so difficult to understand and adjust.

For example, many different factors influence the expression of a gene. Transcription factors bind to nearby DNA, raising or lowering gene expression. Distant sites in the genome act as enhancers or suppressors. DNA winds around histone proteins, in which both the histones themselves and the DNA winding affect expression. The DNA is marked with methyl or acetyl groups, altering expression. A variety of RNAs encoded in other parts of the genome influence different steps in the DNA to RNA to protein process. Why is it all so complex?

The paradox of robustness naturally leads to additional higher level regulatory controls that cause evolutionary relaxation of lower level controls. The complexification is typically irreversible. Additional layers of robustness get added, leading to a deeply and densely wired control system.

Constructive neutral evolution would lead to a similar interpretation. A primary goal of this article is to consider how the paradox of robustness and constructive neutral evolution provide complementary perspectives on complexity, each theory emphasizing different aspects of evolutionary process. Bringing together those alternative perspectives leads to a broader and more powerful framework for understanding robustness and complexity.

\section{Hourglass patterns of robustness and complexity}

This section introduces two patterns of complexity in robust systems. The following section joins those observed patterns with the previously described theory to formulate the broader conceptual framework for future work.

The first pattern concerns the hourglass model of development \autocite{shook08morphogenic,irie14the-developmental}. When comparing related species, the early stages of development tend to diverge relatively rapidly. The intermediate stages diverge relatively slowly, implying stronger conservation or constraint for those stages. The late stages of development diverge relatively rapidly.

Visually, we may think of the early stages as the widely divergent bottom of the hourglass. The intermediate stages are the constrained narrow neck of the hourglass. The late stages of final adult form set the widely divergent top.

A recent study of the nematode \textit{Caenorhabditis elegans} provides an example \autocite{ma23transcriptome}. Proteins that affect early stages of development have evolved relatively rapidly when compared to related species. Proteins that affect intermediate stages have evolved relatively slowly. Proteins that affect late stages have evolved relatively rapidly. The authors interpret this pattern in terms of the classic hourglass model.

The second example concerns the hourglass pattern of design for robust and complex systems in both engineering and biology \autocite{doyle11architecture}. These ideas about robust system architecture come from John Doyle, a major contributor to robust control theory in engineering \autocite{doyle09feedback,zhou98essentials}.

Modern mobile phones provide an example of Doyle's hourglass architecture. Mobile phones are essentially small computers. The hardware aspect of a computer provides a few basic functions \autocite{conrad85on-design,eigenmann98von-neumann,arikpo07von-neumann}. Information needs to be stored in a retrievable way. Digital logic supports programming. Different hardware can offer these same basic functions. 

Various companies manufacture mobile phones. Their hardware designs differ but remain qualitatively equivalent with regard to computation. In Doyle's hourglass, lower-level hardware diversity arises because there are many approximately equivalent ways to provide a basic foundation for similar functions. The lower part of the hourglass is diverse and wide.

The different hardwares are functionally equivalent because they all support the same basic set of protocols. The protocols are the core part of the operating system that sits atop the hardware. An operating system is like Microsoft Windows, which runs on many personal computers, or Mac OS, which runs on Apple computers. Essentially all modern mobile phones run variants of the Linux operating system.

At the base layer of Linux, the kernel sits just above the hardware. When a software program running on a phone needs to store information, it tells the kernel's protocols to store the information. The software program does not know anything about how the hardware actually stores the data. The software only knows how to talk the core protocols. Similarly, the hardware does not know anything about the software layer. The hardware only provides the basis for the core protocols.

The core protocols are highly constrained by the need to provide the common foundation for computation. They do not differ very much from one phone to another, apart from the need to translate messages from the software layer to any special hardware that a phone might have. In Doyle's hourglass, the protocols form the narrow middle waist. The common protocols allow different hardwares to be functionally equivalent, releasing constraint on hardware design and leading to hardware diversity.

The upper software layer creates the functions that make mobile phones useful. The same software can in principle run on any hardware because the software talks only to the commonly used operating system protocols. The upper software layer diversifies widely to match the wide range of functions that users demand. The diverse software layer forms Doyle's wide upper half of the hourglass.

In practice, different manufacturers add an additional software layer between the core operating system protocols and the functional software programs. That intermediate software layer differentiates the upper-level software that can run on the phones of different manufacturers. However, that limitation mostly arises from proprietary business practices rather than from fundamental aspects of engineering design.

At the engineering level, hardware diversifies because there are many physical ways to make a base system layer that supports common protocols. On top of those common protocols, many different functional or software processes can be developed, each talking to the same small set of common protocols. The core protocols act as a buffering layer that deconstrains the need to match hardware and software levels, allowing those levels to evolve in nearly independent ways.

Doyle suggests that essentially all robust complex systems have a similar hourglass architecture, from airplanes to automobiles to communication systems. Csete and Doyle have also argued that robust complex systems in biology have a similar architecture \autocite{csete02reverse,csete04bow-ties}. Consider two examples.

First, essentially all cells of life power themselves by a disequilibrium between ATP and ADP molecules. Roughly speaking, food is used to drive reactions that add a phosphate group to ADP, making ATP. The ATP/ADP disequilibrium acts like a storage battery that provides power to the cell, driving processes that make the biomolecular structures of life and powering functional and behavioral activities. 

Across life, widely conserved biochemical mechanisms create and control the ATP/ADP disequilibrium. Those conserved mechanisms form the core protocols of power at the hourglass's central waist. At a lower, hardware-like level, many different biochemical reactions acquire and process diverse kinds of food. Some organisms can live on methane. Others need sugar. At the wide bottom of the hourglass, diverse biochemistry does the initial processing of various food sources.

As the metabolic cascade moves upward from the diverse initial inputs toward the central ATP/ADP power protocols, the biochemistry narrows to an evolutionarily conserved core. From that narrow core, once the ATP/ADP disequilibrium is in place to provide power, life diversifies widely into different software-like functional programs. Different organisms use that core power to build different kinds of molecules and to function in different ways. The top of the hourglass widens.

Second, essentially all cells of life use the DNA makes RNA makes protein cascade to translate stored hereditary information into the proteins that provide function. Widely conserved protocols process this essential translation. Variations occur but remain tightly constrained by the basic need to use the information in nucleotides to make functional amino acid sequences.

Diversity in genomic information storage, transmission, and retrieval forms a wide hardware-like lower level that flows into the narrow mid-level core protocols. The proteins that emerge from that mid-level build a widely divergent upper level of software-like function. The genome hardware level and the protein software level may diverge broadly and in mostly uncoupled ways.

A recent study by Michael Levin's group suggests a link between the hourglass model of development and Doyle's hourglass model of robust complexity \autocite{shreesha23cellular}. Their computer simulation followed the evolutionary processes that shape development. Genomes encoded only simple rules of cellular processes rather than final developed forms. To solve particular developmental challenges, the genomes evolved to encode a few core protocols of cell-cell interactions and some specifications for how those core protocols were to be used.

The authors interpreted their results in terms of an hourglass model of development: ``[M]utations resulting in noise or changes in initial positions of the organs\dots will not have a strong effect on survival because the competency of the tissues will make needed reconfigurations to compensate for errors in initial state.''

Put another way, the narrowly conserved intermediate developmental stages robustly buffer fluctuations in early developmental steps, leading to the evolutionary diversification of those early steps. They also found that final forms could diverge widely, tracing the classic hourglass shape.

The authors neither used the term ``hourglass'' nor connected their work to those classic theories for the evolution of development. Instead, they rediscovered the hourglass pattern directly from observing how their computer simulations evolved.

Similarly, they rediscovered aspects of the hourglass model for the architecture of robust and complex systems without awareness of Doyle's work. In particular, they emphasized that the genomic hardware evolved nearly independently from the developmental software because the relatively conserved core developmental protocols screened off changes between the hardware and software layers. They emphasized the words ``hardware'' and ``software'' in their interpretations. They also noted that in planaria, a kind of flatworm, genomic changes are often uncoupled from developmental changes.

This study's rediscovery of the hourglass models of development and robust complexity provides a compelling signal. When different investigators start from distinct backgrounds and focal questions and then converge on similar concepts, it often means that the time is right for a new synthesis.

\section{A broader conceptual foundation}

Constructive neutral evolution and the paradox of robustness describe similar processes. A system's higher-level mechanism suppresses the consequences of variability at lower component levels. The components become more variable, perhaps drifting neutrally or evolving to sloppier, lower cost and lower performance states. 

The two theories, although similar, emphasize different aspects of biological design.

Constructive neutral evolution focuses primarily on genomic complexity. That complexity in the storage and transmission of information links to Doyle's hardware level of robust and complex systems. There are many physical ways to manage information. Diversity ultimately matters little as long as the physical variety flows through the common protocol of DNA makes RNA makes protein.

The paradox of robustness focuses primarily on functional complexity. This theory, initially motivated by the variety of component systems that protect against cancer, emphasizes physiological homeostasis, repair of cellular damage, cell suicide to avoid harm, excess capacity to mitigate exceptional challenge, and plasticity and behavioral adjustment to changing environments. These functional protections link to Doyle's software level of robust and complex systems.

At the genomic hardware level, constructive neutral evolution emphasizes how buffering mechanisms often induce neutrality and evolutionary drift in the processes that manipulate information. The particular ways in which information gets processed may not matter so much as long as the information retains the ability to encode proteins. Systems tolerate low-level physical variety that retains support for the essential protocols at the hourglass's narrow waist.

At the functional software level, the paradox of robustness emphasizes how buffering mechanisms often alter the marginal costs and benefits of functional components. For example, in the data storage RAID example, the higher-level RAID controller buffers the consequences of failure at the component hard drive level. Thus, the hard drives became cheaper and sloppier, decaying in marginal cost and benefit to a lower-performing state. The hard drives did not drift neutrally. Rather, they follow economic principles. Robustness at the functional level will often alter economic costs and benefits rather than induce neutrality.

Pushing the analogies, the earliest stages of development create the first physical pieces needed to build an organism. Those physical pieces of hardware can be made in a variety of ways, as long as the basic pieces come into place. Then, at the intermediate stage of development, the hardware pieces have to be organized through the common protocols that shape tissues. Those protocols robustly buffer early variety and provide the functional basis for the software programming that makes diverse adult forms.

Of course, the analogies are far from perfect. But they do seem to capture fundamental aspects of biological design. They also match common patterns in human-engineered systems. 

Previously, the various theories followed isolated lines of thought. The paradox of robustness, constructive neutral evolution, Lindquist's cellular buffering, the hourglass model of development, and Doyle's hourglass model of robust and complex systems arose separately and remained alone. The fact that these ideas fit together in a natural and cohesive way suggests progress toward a comprehensive foundation for understanding biological design. 

\section{Ratchet of complexity: evolutionary consequences}

The more seemingly separate problems that fit into our new framework, the more evidence we have of moving in the right direction. This final section considers one further step toward conceptual unification. Can we link our broad framing for the biological evolution of robust and complex systems to recent progress in machine learning and artificial intelligence? Biological evolution is a particular kind of learning process. Links with machine learning would not be surprising.

Recently, deep computational neural networks provided several breakthroughs in applications \autocite{goodfellow16deep}. Part of the success came from using deep multilayer networks that are densely connected. The huge parameter space of these models typically overfits the data. In spite of that overfitting, the models often generalize well, with excellent performance on test data not used in the fitting process. This benign overfitting remains an unsolved puzzle \autocite{bartlett20benign,elad20another,papyan20prevalence}.

The paradox of robustness creates deeply densely connected networks in evolutionary systems. With each addition of robustness at the system level, the lower-level components relax evolutionarily, causing some decay. Subsequently, the system cannot reverse by removing the new robustness mechanism because the decayed components would perform poorly when not protected by the additional robustness. An irreversible layer of complexity has been added.

Eventually, a new high-level robustness mechanism may be favored, layered above the existing system. The process repeats, with decay of lower-level components and an irreversible ratchet of increasing complexity \autocite{gray10irremediable,frank07maladaptation}. Eventually the system becomes a deeply layered and densely wired architecture \autocite{frank17puzzles5}. If deeply densely wired systems do in fact learn particularly well, then such overparameterized evolutionary systems may adapt particularly rapidly and effectively to novel challenges. Perhaps life owes part of its great evolutionary success to the inevitable overwiring that arises from the paradox of robustness.

Evolution proceeds by incremental trial and error. Other kinds of systems designed by incremental trial and error may share similar features. Human institutions come to mind. Incremental changes may be more common than global redesign. If so, we may expect that system-wide error correction leads to the decaying performance of subunits, a layered architecture, and irreversible complexity.

\section*{Acknowledgments}

\noindent The Donald Bren Foundation, National Science Foundation grant DEB-1939423, and DoD grant W911NF2010227 support my research. This manuscript arose from a prior video presentation available at \href{https://youtu.be/LP1-vQ3zYgM}{https://youtu.be/LP1-vQ3zYgM}.

%\vfill\eject

\mybiblio	% uses main.bib by default, add other bibs as needed

% used cuted package strip env to force balancing of columns
\ifmulticol\begin{strip}\hbox{\null}\end{strip}\hbox{\null}\fi

\end{document}